\documentstyle{article}
\begin{document}
\LARGE
\begin{center}
\bf Quantum Fields in Schwarzschild-de Sitter Space
\vspace*{0.4in}
\normalsize \large \rm 

Wu Zhong Chao

International Center for Relativistic Astrophysics

Rome University, Italy

and

Dept. of Physics

Beijing Normal University

Beijing 100875, China

\vspace*{0.3in}
\large
\bf
Abstract
\end{center}
\vspace*{.1in}
\rm
\normalsize

In the No-Boundary Universe a primordial black hole is created
from a constrained 
gravitational instanton. The black hole created is immersed in
the de Sitter background
with a positive gravitational constant. The constrained instanton
is characterized not only by the external parameters, the mass
parameter, charge and angular momentum, but also by one more
internal parameter, the identification period in the
imaginary time coordinate. 
Although the period has no effect on the black hole
background, its inverse is the 
temperature of the no-boundary state of the perturbation modes
perceived by an observer. By using the Bogoliubov transformation,
we show that the perturbation modes of both scalar and spinor
fields are in thermal
equilibrium with the black hole background at the arbitrary
temperature. However, for the two extreme cases, the 
de Sitter and the Nariai models, the no-boundary  state remains 
pure.

\vspace*{0.3in}

PACS number(s): 98.80.Hw, 98.80.Bp, 04.60.Kz, 04.70.Dy


e-mail: wu@axp3g9.icra.it 

\pagebreak

\large \bf I. Introduction
\vspace*{0.15in}

\rm 

\normalsize

The First Cause Problem has been dispelled by the Hawking theory
of quantum cosmology. The no-boundary proposal, in
principle, has the power to predict the evolution of the universe
and its matter content [1]. However, due to the technical
difficulty, the calculation has
only been carried out for some special models. At this moment it
is believed that the Hawking massive scalar model is the most
realistic one [2]. In this model, the evolution of the early
stage
in real time can be approximated by a de Sitter spacetime
with an effective cosmological constant $\Lambda$. A de Sitter
universe can be thought of as originating through a quantum
transition from a Euclidean $S^4$ space. Since the $S^4$ space is
the most symmetric  manifold, it is conjectured that the de
Sitter expansion is the most probable evolution in the Planckian
era of the universe.

On the other hand, a black hole is the simplest, i.e. the most
beautiful object except for the universe itself in science. The
discovery of Hawking radiation is the most dramatic event in
gravitational physics for several decades [3]. Unfortunately, the
time scale of evaporation for a macroscopic black hole is much
longer than the age of the universe. The only hope
to observe the radiation is from those microscopic black holes
formed in the very early universe. These are so-called primordial
black holes. A gravitational collapse of a matter fluctuation can
lead to the black hole formation. However, in this paper we shall
only consider the most interesting and dramatic way, that is, the
black hole is quantum mechanically created at the same moment
of the birth of the universe. If the evolution of the very early
universe is dominated by the presence of the effective
cosmological constant, then the primordial black hole should
be immersed in the background of the de Sitter spacetime. Its
metric can be described by a member of the Kerr-Newman-de Sitter
family [4].

In quantum cosmology it was thought that, at the $WKB$ level, the
Lorentzian evolution could be obtained through an analytic
continuation from a instanton  at its
equator $\Sigma$. This is true only for some special cases, like
the de Sitter model and the Nariai model which represents a pair
of black holes. Here, the black hole mass parameter takes two
discrete values $0$ and $m_c = \Lambda^{-1/2}/3$, and the 
charge $Q$ and angular momentum $ma$ are  vanishing [5]. The
quantum
creations of these two models have maximum and minimum
probabilities, respectively, in comparison with black hole
creation with different  parameters $m$, $Q$ and $ma$ and the
same
cosmological constant. In general, if a wave function represents
an ensemble of evolutions with  continuous parameters, like the
case of a primordial black hole, then it is unlikely that all
these trajectories of the ensemble can be obtained through an
analytic
continuation from an ensemble of instantons. Unless the action is
a constant function of the parameters, then the action cannot be
stationary with respect to them in the range of the parameters.
On the other hand, the condition
for the existence of an instanton is that its action must be
stationary.

It was realized recently that a Lorentzian evolution
does not need to  originate from a regular instanton. Instead,
a generalized version of an instanton can provide a seed for the
evolution. When one calculates the creation probability 
through a path
integral, the sum is over all no-boundary compact 
4-metrics with the given 
$\Sigma$ as its sector of the quantum transition. The dominating 
contribution is due to the
manifold with a stationary action. Consequently, the manifold
should 
satisfy the Einstein and other field equations everywhere with
the possible
exception of the equator $\Sigma$. This is derived from the
variational calculation with the constraints imposed by the 
3-metric and matter fields on $\Sigma$. The stationary action
solution 
is called a constrained gravitational instanton [6].

This is exactly what happens in the case of primordial black
hole creation. Here, the topology of the space is $S^2 \times
S^1$,
instead of $S^3$ in the de Sitter model. The constrained
instanton is obtained through a periodic identification of the
imaginary time in the Euclidean solution. No period can make the
whole manifold regular except for the de Sitter model, the Nariai
model
or some special models. The identification
will lead to at least one conical singularity at the black hole
and cosmological horizons. It turns out that although the
Euclidean action is not stationary with respect to the three
parameters $m$, $Q$ and $ma$ which characterize the 3-metric and
matter fields
on $\Sigma$, it is independent of the period. Thus  the action is
stationary under the
constraint imposed by the configuration of the wave function at
the quantum transition surface. Therefore, we have found the
constrained gravitational instanton [6].

The equator passes through the two
horizons of the instanton. The Lorentzian evolution emanates from
the equator
regardless of the identification period. Furthermore,
since at the $WKB$ level,  the probability is the exponential of 
the negative of the
instanton action, and we know the action
 is independent of the period, it follows that there
is no effect of the period on the background of spacetime. This 
is in sharp contrast with the case of a back hole immersed in an
asymptotically flat space. The instanton action of a
asymptotically flat
black hole does depend on the period. The probability of our
black
hole  creation is the  exponential of a quarter of
the sum of the black hole and cosmological horizons, or the
exponential of the total entropy of the universe. The probability
is a decreasing function of the magnitudes of these parameters.
Therefore,
our result supports the claim that the de Sitter evolution is
most probable [6].

On the other hand, if we study the quantum state of perturbation
modes
in the black hole background, then the no-boundary proposal
implies
that it should be in the minimum excitation state. However, an
observer
can only measure the configuration within his black hole and
cosmological
horizons. One has to take the trace of the wave function over the
unobserved configuration  of the perturbation mode. 
The quantum state observed is described by a density
matrix. This matrix represents a thermal equilibrium state with a
temperature represented
by the inverse of the identification period in imaginary time.
The arbitrariness 
of the period means that the fields can be in thermal 
equilibrium with the black hole at any temperature.

It is interesting to note that in the de Sitter model, the black
hole horizon disappears,
while in the Nariai model, the two horizons degenerate. For both
cases, nothing prevents an observer from measuring the quantum
state at the whole equator. Therefore the associated temperature
is  zero. The arbitrariness of the identification period has
effect neither on the background spacetime nor on the
perturbation modes.

At the $WKB$ level, the Wheeler-DeWitt equation in quantum
cosmology can be decomposed into two parts. One is for the
background, the other is for the perturbation modes, if we ignore
the back reaction of the perturbations to the background. Then a
physical time coordinate naturally emerges from the wave packet
for the background. The second part of the Wheeler-DeWitt
equation takes the form of the Schroedinger equation in the
spacetime background. One shall work in the framework of quantum
fields in curved spacetime supplemented by the no-boundary
condition.

We shall consider a massive scalar field and a massive spin-1/2
field in
the Schwarzschild-de Sitter background. By using the Bogoliubov
transformation, we show that to an observer whose observation is
always restricted by the two horizons, the no-boundary quantum
state of bosonic or fermionic fields presents  a thermal
equilibrium state with temperature being the inverse of the
imaginary time period. Therefore, in the Schwarzschild-de Sitter
spacetime, the quantum fields can coexist with the black hole at
an equilibrium state with an arbitrary temperature. In the two
extreme cases, the de Sitter model and the Nariai model, since
the topologies of the equators are different from the nonextreme
case, nothing prevents the observer from measuring the fields on
the whole equators and the quantum states remain pure.

We shall review the approach of constrained gravitational
instanton in Sec. II.  This method will be used in Sec. III to
study the quantum state of the  Kerr-Newman-de Sitter metric
family, which represents a  single primordial  black hole
immersed
in de Sitter spacetime background. Sec. IV investigates the
no-boundary quantum state of the scalar field in the 
background. A similar calculation is carried out for the spinor
field in Sec. V.  Sec. VI is a discussion of this.

\vspace*{0.3in}

\large \bf II. The constrained gravitational instanton

\vspace*{0.15in}

\rm 
\normalsize
In the No-Boundary Universe, the wave function of the
universe is given by the Hartle-Hawking ground  state [1]
\begin{equation}
\Psi(h_{ij}, \phi) = \int_{C}d[g_{\mu \nu}] d [\phi] \exp (-
\bar{I} ([g_{\mu \nu}, \phi]),
\end{equation}
where the path integral is over the class $C$ of  all compact
Euclidean $4$-metrics and matter field configurations which
agree with the given $3$-metrics $h_{ij}$ of the only boundary
and matter configuration $\phi$ on it. Here $\bar{I}$ means the
Euclidean action.

The Euclidean action for the gravitational part of a smooth
spacetime manifold $M$ with boundary $\partial M$ is
\begin{equation}
\bar{I} = - \frac{1}{16\pi} \int_M (R - 2\Lambda) -
\frac{1}{8\pi} \int_{\partial M} K,
\end{equation}
where $\Lambda$ is the cosmological constant, $R$ is the scalar
curvature and $K$ is the trace of the second fundamental form of
the boundary.

The dominant contribution to the path integral comes from some
stationary action manifolds with matter fields on them, which are
the saddle points for the path integral. Then the wave 
function takes a superposition form of wave packets
\begin{equation}
\Psi \approx C \exp (- S/\hbar),
\end{equation}
where we have written $\hbar$ explicitly; $C$ is a slowly varying
prefactor; and $S \equiv S_r + iS_i$ is a complex phase. $S$ is
identified with the action $\bar{I}$ here.

Since the wave packets of form (3) are not independent in the
decomposition of the wave function, one more restriction should
be imposed.  That is,  the wave packets themselves should obey
the
Wheeler-DeWitt equation. Classically, this means that the
evolutions represented by the wave packet should satisfy the
Einstein equation with some quantum corrections, as will be
shown below.

From the Hartle-Hawking proposal one can derive the probability
of the 3-surface $\Sigma$ with the matter field $\psi$ on it 
\begin{equation}
P = \Psi^\star \Psi = \int_C d [g_{\mu \nu}] d[\psi] \exp(-
\bar{I} ([g_{\mu \nu}, \psi ]),
\end{equation}
where the class $C$ is composed of all no-boundary compact
Euclidean 4-metrics
and matter field configurations which agree with the given
3-metric $h_{ij}$ and the matter field $\psi$ on $\Sigma$. We
shall concentrate on the 3-geometries at which  a quantum
transition from a so-called Euclidean sector to a Lorentzian
sector occurs at the $WKB$ level. These two sectors are
represented by the wave packets with purely real and imaginary
phases. These kinds of transitions are quite special and are
called
real tunneling.  

Here, we do not restrict the class $C$ to regular metrics
only, since the derivation from Eq. (1) to Eq. (4) has already
led to some jump discontinuities in the extrinsic curvature at
$\Sigma$.

The main contribution to the path integral in Eq. (4) comes from
the stationary action 4-metric, which meets all requirements on
the 3-surface $\Sigma$. At the $WKB$ level,
the exponential of the negative of the stationary action is the
probability of the corresponding Lorentzian trajectory.

From the above viewpoint, extending  the class $C$ to
include
metrics with some mild singularities is essential. Indeed, it is
recognized that,
 in some sense, the set of all regular metrics is not complete. 
For many cases, under the usual regularity conditions and the
requirements at the equator $\Sigma$, there may not exist any
stationary action metric, i.e. gravitational instanton.
Therefore, it seems reasonable to include metrics
with jump discontinuities of extrinsic curvature and their
degenerate cases, that is, the conical or pancake singularities
[7].

If we lift the requirement on the 3-metric of the equator, then
the stationary action solution becomes the regular gravitational
instanton, as it satisfies the Einstein equation everywhere. Then
the following Gibbons-Hartle condition should hold at the equator
[8]
\begin{equation}
K_{ij} = 0,
\end{equation}
where $K_{ij}$ is the extrinsic curvature of the equator. The
probability of the corresponding trajectory takes stationary
value; it may be maximum, minimum or neither [9].

If the regular gravitational instanton has minimum action, then
the Lorentzian evolution emanating from it is singled out as the
most probable trajectory. Therefore, quantum
cosmology fully realizes its prediction power: there is no degree
of freedom left [9]. Without using the
instanton theory, the degree of freedom is only reduced to half 
by the ground state proposal. Roughly speaking, this is due to
the
regularity condition at the south pole of the Euclidean manifold
in the path integral.

If in addition, one needs to calculate the probability of a
trajectory emanating from a given 3-metric, then the extra
requirement  on $\Sigma$ is so strong that no gravitational
instanton exists, unless the 3-metric coincides with a sector
of a regular gravitational instanton. However, one can still find
a stationary action trajectory from variational calculus.
From the variational principle, the trajectory obeys the Einstein
equation and other field equations on the 4-manifold with some
exceptions on $\Sigma$, and the Gibbons-Hartle condition is no
longer valid there.

In this way one can find a irregular gravitational instanton
with some mild singularities within the class $C$. It is not a
gravitational instanton in the conventional sense, and is called
a constrained gravitational instanton. Even if the action is
stationary under the constraint of $\Sigma$, the action or
the probability
of the real tunneling associated with it will not be stationary
when
lifting the 3-metric requirements. This is in comparison within
the set of all classical trajectories.

In summary, the action of a constrained instanton is not
necessarily
stationary with respect to the so-called external parameters
which characterize the  3-metric of quantum transition and matter
field on it unless one seeks the most probable Lorentzian
trajectory.  However, the action should be stationary with
respect
to the so-called internal parameters which are the rest of the
continuous parameters. In fact, if there exist some internal
parameters, the action should be independent of them. The reason
is as follows: if the action depends on some of the internal
parameters, then the instantons can be singled out from them by
using the stationary action condition, and
these parameters are not qualified and should be excluded from
the very beginning.

The occurrence of singularities here can be considered as a
purely quantum effect at the $WKB$ level. One should welcome the
situation, if he admits that Nature is quantum. Classically, 
one can  say that the solution obeys the generalized Einstein
equation in the whole manifold.

\vspace*{0.3in}

\large \bf III. The quantum state of the spacetime background

\vspace*{0.15in}

\rm 
\normalsize

It is believed that the universe at the Euclidean and inflation
stage can be approximated by an $S^4$ space and a de Sitter
space with an effective cosmological constant $\Lambda$. In the
Hawking massive scalar model [2], one can set $\Lambda = 
 3m^2_0 \varphi^2_0$, where $\varphi_0$ is the initial 
value of the scalar field. A primordial black hole is sitting in
the background of the de Sitter spacetime. 
A chargeless and non-rotating black hole with the de Sitter
background can be described by the Schwarzschild-de Sitter
spacetime. It is the unique spherically symmetric vacuum solution
to the Einstein equation with a cosmological constant $\Lambda$.
The $S^2 \times S^2$ Nariai spacetime is its degenerate case, and
it represents a pair of black holes.

Its Euclidean metric can be written as 
\begin{equation}
ds^2 = \left (1- \frac{2m}{r} - \frac{\Lambda r^2}{3} \right
)d\tau^2 
+\left (1- \frac{2m}{r} - \frac{\Lambda r^2}{3}\right )^{-1}dr^2 
+ r^2 d\Omega^2_2. \;\;\; (\tau = it)
\end{equation}

The black hole and cosmological horizons $r_2$ and $r_3$ are
determined by the factorization of the potential 
\begin{equation}
\Delta = 1- \frac{2m}{r} - \frac{\Lambda r^2}{3}  = -
\frac{\Lambda}{3r}
(r - r_0)(r - r_2)(r - r_3),
\end{equation}
where $r_0$ is the horizon for the negative $r$. We are
interested in the Euclidean sector $r_2 \leq r \leq r_3$ for $0
\leq m \leq m_c = \Lambda^{-1/2}/3$. For the extreme case $ m =
m_c$ the sector degenerates into the $S^2 \times S^2$ space, or
the Euclidean version of the Nariai metric [5]. 
 
The surface gravities $\kappa_2$ and $\kappa_3$ on the two
horizons are [4]
\begin{equation}
\kappa_2 = \frac{\Lambda}{6r_2}(r_3 - r_2)(r_2 - r_0),
\end{equation}
\begin{equation}
\kappa_3 = \frac{\Lambda}{6r_3}(r_3 - r_2)(r_3 - r_0).
\end{equation}

A constrained gravitational instanton can be constructed as
follows [6].
In the $(\tau - r)$ plane $r = r_2$ is an axis of symmetry,
the imaginary time coordinate $\tau$ is identified with period
$\beta_2 = 2\pi \kappa_2^{-1}$, and $\beta_2^{-1}$ is the
Hawking temperature. This makes the Euclidean manifold
regular at the black hole horizon. One can also apply this
procedure to the cosmological horizon with period $\beta_3 = 2\pi
\kappa_3^{-1}$, and $\beta^{-1}_3$ is the Gibbons-Hawking
temperature [4]. For the $S^2 \times S^2$ case, these
two horizons are identical, and one obtains a regular
instanton. Except for the $S^2 \times S^2$ spacetime, one cannot
simultaneously regularize the whole manifold at both horizons due
to the inequality
of the two temperatures.

One can make two cuts along $\tau = consts.$ between $r = r_2$
and
$r = r_3$ and then identify them. Then a $f_2$-fold cover turns
the $(\tau - r)$ plane into a cone with a deficit angle of
$2\pi (1-f_2)$ at the black hole horizon. In a similar way, one
can have an $f_3$-fold cover
at the cosmological horizon. To construct a symmetric
manifold, $f_2$ and $f_3$ can be any
pair of real numbers satisfying the relation 
\begin{equation}
f_2 \beta_2 = f_3 \beta_3.
\end{equation}

Consequently, the parameter
$f_2$ or $f_3$ is the only degree of freedom left.  One only
needs to see whether the above action is stationary with respect
to this parameter in order to make sure the manifold obtained is
truly a constrained gravitational instanton. We set 
$\Delta \tau = |f_2 \beta_2|$ below. If $f_2$ or $f_3$ is
different from $1$,
then the cone at the black hole or cosmological horizon will have
an extra contribution to the action. The integral of $K$ with
respect to the $3$-area in the
boundary term of the action (2) is the area increase rate along
its normal. Thus, the extra contribution due to the conical
singularities can be considered as the degenerate form 
\begin{equation}
\bar{I}_{2,deficit} = - \frac{1}{8 \pi}\cdot 4\pi r_2^2\cdot 2\pi
(1 - f_2),
\end{equation}
\begin{equation}
\bar{I}_{3,deficit} = - \frac{1}{8 \pi}\cdot 4\pi r_3^2\cdot 2\pi
(1 - f_3).
\end{equation}
The addition of the boundary term to the volume action is
equivalent to the change from the extrinsic curvature
representation to the
metric representation. For our case, at the classical level, the
momentum at the equator vanishes except for the two horizons. At
the horizons, the pair of canonical conjugate variables can be
described by the horizon area and the deficit angle. Taking into
account the fact that the extrinsic curvature vanishes elsewhere,
the terms $\bar{I}_{i,deficit}$ are the only nonvanishing
Legendre terms.

The volume term of the action for the manifold can be calculated
as
\begin{equation}
\bar{I}_{vol} = -\frac{\Lambda}{6} (r^3_3 - r^3_2) f_2 \beta_2.
\end{equation}

Using Eqs. (10) - (13), one can get the total action
\begin{equation}
\bar{I}_{total} = - \pi (r^2_2 + r^2_3).
\end{equation}
This is one quarter of  the negative of the sum of the two
horizon areas.

One readily notices that the action is independent of the
choice of $f_2$ or $f_3$.  Our result
(14) shows that the constructed manifold is indeed a constrained
gravitational instanton, and $f_2$ or $f_3$ is identified as
the internal parameter. We can set $\tau = \pm \Delta \tau /4 $
 for the equator of the quantum transition. Our calculation shows
that no matter which flat fragment
of the constrained gravitational instanton is chosen, the same
black hole should be created with the same probability. Of
course, the most dramatic case is that of no volume, i.e. $f_2 =
f_3 = 0$. 

The probability of the black hole creation is 
\begin{equation}
P_m \approx \exp (\pi(r_2^2 + r_3^2)).
\end{equation}

This result interposes  two special cases [5]. The first is the
de Sitter
model with $m = 0$,
\begin{equation}
P_0 \approx \exp \left ( \frac{3\pi}{\Lambda} \right )
\end{equation}
and the second is the Nariai model,  or pair black hole creation,
with
$m = m_c$,
\begin{equation}
P_c \approx \exp \left ( \frac{2\pi}{\Lambda} \right ).
\end{equation}

The probability is an exponentially decreasing function in terms
of the mass parameter. The de Sitter case has the maximum 
probability and the Nariai case has the minimum probability.

If one includes an electromagnetic field into the model, one
would 
be able to carry out a similar calculation [10]. One just simply
replaces 
the potential by
\begin{equation}
\Delta = 1 - \frac{2m}{r} + \frac{Q^2}{r^2} - \frac{\Lambda
r^2}{3}
= - \frac{\Lambda}{3r^2}(r - r_0)(r - r_1)(r - r_2)(r - r_3),
\end{equation}
where $Q$ is the charge parameter of the black hole. 

For the magnetically charged black hole case, the configuration
of the
wave function is the 3-metric and magnetic charge. However,
the configuration for the wave function of an electrically
charged 
black hole is not well defined [5][6][11][12], if one naively
uses the folding 
and gluing techniques described above. For the electric case, the
configuration of the wave function is the 3-metric and 
the canonical momentum conjugate to the charge. 
In order to get the wave function for the charge, one has to
appeal to a Fourier transformation, by which the duality between
electric and magnetic black holes is recovered.

One can also investigate the problem of creation of a rotating
black hole, that is,  of a Kerr-Newman-de Sitter black hole. The
configuration of the wave function from the naive cutting,
folding and gluing is the 3-metric, matter field and the
differentiation
rotation of the two horizons. One has to use another Fourier
transformation to obtain the wave function for the angular
momentum. More crucially,  only under this consideration, is the
Euclidean action stationary, and our
construction becomes meaningful. 

The probability of black hole creation is the exponential of a
quarter of both black hole and cosmological horizon areas, or the
exponential of the total entropy. It is an exponentially
decreasing function
with respect to the mass, charge amplitude and angular momentum.
The de Sitter spacetime without black hole is the most probable
evolution at the Planckian era. Due to the No-Hair theorem, the
problem of a single black hole creation in quantum cosmology
has been completely
resolved [6].

In this paper we study quantum fields in the 
Schwarzschild-de Sitter black hole background only. 

The whole scenario of the Schwarzschild-de Sitter black hole
creation is shown in Fig. 1. The $S^2$ space $(\theta-\phi )$ is
represented by a $S^1$ space around the vertical axis. The radius
of $S^2$ is $r$. The bottom part is the instanton, the upper part
shows the black hole created. It shows the collapsing of the
internal edge 
of the doughnut
from the black hole horizon into the
singularity $r = 0$ and the expansion of the external edge from 
the
cosmological horizon to $r
= \infty$. The $S^1$ equator in space $(\tau- r)$ and the conical
singularities are not explicitly shown here. The observer can
only measure the fields at one half of the equator between the
two horizons.

The global aspect of the scenario is depicted by the
Penrose-Carter diagram in
Fig. 2. The $S^2$ space $(\theta - \phi)$ is depressed. There is
an
infinite sequence of diamond shape regions, singularities $r = 0$
and spacelike infinities $r = \infty$. The scenario of the black
hole creation can be obtained by an identification, for instance,
with lines $ABC$ and $DEF$ [13]. The equator in the instanton is
identified as the closed line $BE$ here. We shall set $t = 0$ and
$\tau = \Delta \tau /4$ for line $\Sigma_1$ or $OE$ and $ t =0 $
and
$\tau = -\Delta \tau /4$ for the line $\Sigma_2$ or $BO$.

\vspace*{0.3in}

\large \bf IV. The quantum state of the bosonic field

\vspace*{0.15in}

\rm 
\normalsize

Now we are going to discuss the quantum state of perturbation
modes in the Schwarzschild-de Sitter spacetime
background. The quantum fields in the Schwarzschild spacetime
background has been studied [14][15][16][17][18][19]. 

We shall show that the quantum state of the matter fields in the
same black hole background is characterized by the
parameter of the identification period in the imaginary time.

If we neglect the back reaction of the perturbations to the
background, then  the wave function of the universe for a fixed
$\Delta \tau$, at the $WKB$ level, can be written as a product
[20]
\begin{equation}
\Psi (h_{ij}, \varphi, \psi ) =
 \Psi(h_{ij})\Psi(h_{ij}, \varphi) \Psi(h_{ij},\psi),
\end{equation}
where $ \Psi(h_{ij})$ is the wave function for the spacetime
background, $\Psi(h_{ij}, \varphi)$ is the
wave function for the scalar perturbation modes and $
\Psi(h_{ij}, \psi )$  is the wave function for the spinor
perturbation modes. The contributions to the wave functions come
from the
classical trajectories which satisfy the no-boundary condition.
The internal time coordinate emerges naturally from the wave
packet $\Psi(h_{ij})$ for the background. The Wheeler-DeWitt
equation can be decomposed into several parts. The parts for the
perturbation modes take the form of the 
Schroedinger equation in the spacetime background. It means, at
the level of our approximation, the study of perturbation modes
in quantum cosmology can be reduced to  that of quantum
fields in curved spacetime supplemented by the no-boundary
condition. This is exactly what we are doing now. 
We shall defer the spinor case to the next section.

The action of the  scalar field $\varphi$ is
\begin{equation}
\bar{I}_s = \frac{1}{2}\int_M  (
g^{\mu\nu}\partial_{\mu}\varphi
\partial_{\nu} \varphi + m^2_0 \varphi^2).
\end{equation}

From the action one can derive the Euclidean equation of motion
\begin{equation}
[\Box - m^2_0 ] \varphi = 0.
\end{equation}

Its basis of  solutions are
\begin{equation}
\omega_{\omega l \bar{m} A }= \exp (\mp
\omega \tau)R_{\omega l \bar{m} A }(r, \theta, \phi)\;\;\;\;\;
(\omega > o).
\end{equation}

The eigenfunctions for the scalar field takes the form
\begin{equation}
R_{\omega l \bar{m} A }(r, \theta, \phi) =R_{\omega l A
}(r)Y_{l\bar{m}}
(\theta, \phi),
\end{equation}
where $Y_{l\bar{m}}$ are the real spherical harmonics,
$l$,$\bar{m}$ denote
the angular momentum magnitude and its component along a given
direction, and $A$ is the label of the two independent radial
functions. The radial function satisfies the equation
\begin{equation} 
\left [ \frac{d}{dr} (r^2 - 2mr + \frac{\Lambda r^4}{3})
\frac{d}{dr} - l(l+1) + \omega^2 \frac{r^2}{1 - \frac{2m}{r} +
\frac{\Lambda r^2}{3}}- m_0^2 r^2 \right ] R_{\omega l A} (r) =
0.
\end{equation}

The eigenfunctions satisfy the orthonormality and completeness
conditions in the space $(r, \theta, \phi)$ with the weight
factor $g^{\tau \tau}g^{1/2}$.

If one rescales the radial function by $\bar{R}_{\omega l A }(r)
=
\sqrt{2\pi}rR_{\omega l A }(r)$, then the equation takes the form
\begin{equation}
\frac{d^2 \bar{R}_{\omega l A }}{dr^{*2}} +  (\omega^2 -
V_l)\bar{R}_{\omega l A } = 0,
\end{equation}
where
\begin{equation}
r^* = \frac{3}{\Lambda} \left (\frac{r_0\ln
|r - r_0|}{(r_0-r_2)(r_0-r_3)} +\frac{r_2\ln |r - r_2|}{(r_2
-r_0)(r_2-r_3)}
+\frac{r_3\ln |r - r_3|}{(r_3 -r_2)(r_3-r_0)} \right )
\end{equation}
and 
\begin{equation}
V_l = \left (1 - \frac{2m}{r} + \frac{\Lambda r^2}{3} \right )
\left ( \frac{l(l+1)}{r^2} + \frac{2m}{r^3} + \frac{2\Lambda}{3}
+ m^2_0 \right ).
\end{equation}

For the given scalar field $\varphi_{\pm}(r, \theta, \phi)$ at
the boundary $\Sigma$,  
the scalar solution can be written as a linear combination
\begin{equation}
\varphi (\tau, r, \theta, \phi) =\sum_\lambda [\varphi_{\lambda,
+}
u_{\lambda, -}(\tau, r_+, \theta, \phi)+\varphi_{\lambda, -}
u_{\lambda, +}(\tau, r_-, \theta, \phi)],
\end{equation}
where $r_+$ and $r_-$ are to distinguish the $r$ coordinate at
$\Sigma_1$ and $\Sigma_2$, and the basis functions are
\begin{equation}
u_{\lambda, \pm}(\tau, r, \theta, \phi)
= \frac{\sinh [(\Delta \tau /4 \mp \tau)\omega]}{ \sinh (\Delta
\tau \omega /2)}R_{\omega l \bar{m} A }(r, \theta, \phi),  
\end{equation}
\begin{equation}
\lambda = (\omega, l, \bar{m}, A), \;\; \sum_\lambda =
\frac{1}{\sqrt{2\pi}} \int^\infty_0 d \omega \sum_{l. \bar{m} A}.
\end{equation}
and the coefficients $\varphi_{\lambda,\pm}$ are determined by
the decomposition  at the boundary $\Sigma$
\begin{equation}
\varphi_{\pm} (r, \theta, \phi) = \sum_\lambda
\varphi_{\lambda,\pm}R_\lambda (r, \theta, \phi).
\end{equation}
We shall assume that the observer is at the cut $\Sigma_1$.

The Euclidean action  can be evaluated by substituting (29) into
(20), yielding
\begin{equation}
\bar{I}_s= \frac{1}{2} \sum_\lambda \left [ \omega_{\lambda}
\coth
(\Delta \tau \omega_{\lambda}/2)(\varphi^2_{\lambda,+}
+\varphi^2_{\lambda,-}) - 2\omega_\lambda \mbox{cosech}
(\Delta \tau
\omega_{\lambda}/2)\varphi_{\lambda,+}\varphi_{\lambda,-} \right
].
\end{equation}

The action can be rewritten into the form  
\begin{equation}
\bar{I}_s= \frac{1}{2} \sum_\lambda \omega_{\lambda}\left [ 
\eta^2_{\lambda, +} +\eta^2_{\lambda, -}\right ],
\end{equation}
where the new variables $\eta_{\lambda, \pm}$ are
defined as follows
\begin{equation}
\varphi_{\lambda,\pm}= \left (2 \sinh \frac{\Delta \tau
\omega_{\lambda}}{2}\right )^{-1/2} \left [e^{\Delta \tau
\omega_\lambda/4}\eta_{\lambda,\pm} + e^{-\Delta \tau
\omega_\lambda/4}\eta_{\lambda,\mp} \right ],
\end{equation}

The no-boundary wave function at the $WKB$ level is due to the
regular solution with the boundary condition (31), it can be
evaluated using the action (33) 
\begin{equation}
\Psi_{NB} (\varphi_+, \varphi_-) \approx  \exp -
\bar{I}_s = \exp \left (-\frac{1}{2} \sum_\lambda
\omega_{\lambda} ( 
\eta^2_{\lambda, +} +\eta^2_{\lambda, -})\right ),
\end{equation}
the wave function takes the Gaussian minimum excitation form.
Thus, if one can simultaneously measure the quantum state within
the whole transition sector, he will find that the scalar field
is in the ground state of these modes, in the sense of usual
quantum mechanics. However, the observation is restricted by the
two horizons and the quantum state will appear to be mixed due
to the information loss as we shall show later.

The operator version of the scalar field $\varphi$ can be
expanded in the Lorentzian spacetime as
\[
\varphi (t, r, \theta, \phi) 
=\sum_{\lambda }\frac{1}{\sqrt{2\omega_\lambda}}
[a_{\lambda,+}u_{\lambda, -}(t, r_+,\theta, \phi)
+a^\dagger_{\lambda, +}u_{\lambda, -}^*(t, r_+, \theta, \phi)
\]
\begin{equation}
+a_{\lambda,-}u_{\lambda, +}(t, r_-, \theta, \phi)
+ a^\dagger_{\lambda,-}u_{\lambda, +}^*(t, r_-, \theta, \phi)].
\end{equation}
where $a_{\lambda, \pm},a_{\lambda, \pm}^\dagger$ are the
annihilation and creation operators associated with the modes
$\varphi_{\lambda, \pm}$ at $\Sigma_1$ and $\Sigma_2$.
Each mode can be considered as a linear oscillator.
Here the modes $u_{\lambda, \mp}$ are not of purely positive or
negative frequency  due to the acausal propagation of the
particles in the spacetime with nontrivial topology.

In the $\varphi_{\lambda, \pm}$ representation, one has [21]
\begin{equation}
a_{\lambda, \pm} +a_{\lambda, \pm}^\dagger =
(2\omega_{\lambda,\pm})^{1/2}\varphi_{\lambda,\pm},
\end{equation}
\begin{equation}
a_{\lambda, \pm} -a_{\lambda, \pm}^\dagger =
\left (\frac{2}{\omega_{\lambda,\pm}}\right
)^{1/2}\frac{1}{\partial
\varphi_{\lambda,\pm}}.
\end{equation}

If one expands the scalar field in terms of the modes
$\eta_{\lambda, \pm}$, then the associated annihilation and
creation operators $c_{\lambda, \pm},c_{\lambda, \pm}^\dagger$
in a similar way satisfy
\begin{equation}
c_{\lambda, \pm} +c_{\lambda, \pm}^\dagger =
(2\omega_{\lambda,\pm})^{1/2}\eta_{\lambda,\pm},
\end{equation}
\begin{equation}
c_{\lambda, \pm} -c_{\lambda, \pm}^\dagger =
\left (\frac{2}{\omega_{\lambda,\pm}}\right
)^{1/2}\frac{1}{\partial
\eta_{\lambda,\pm}}.
\end{equation}

From (34) one can derive
\begin{equation}
\frac{1}{\partial \varphi_{\lambda,\pm}}= \left (2 \sinh
\frac{\Delta \tau
\omega_{\lambda}}{2}\right )^{-1/2} \left [e^{\Delta \tau
\omega_\lambda/4}\frac{1}{\partial \eta_{\lambda,\pm}} -
e^{-\Delta \tau
\omega_\lambda/4}\frac{1}{\partial \eta_{\lambda,\mp}} \right ].
\end{equation}

From Eqs. (34), (37)-(41) it follows that
\begin{equation}
a_{\lambda,\pm} = \left (2 \sinh \frac{\Delta \tau
\omega_{\lambda}}{2}\right )^{-1/2} \left [e^{\Delta \tau
\omega_\lambda/4}c_{\lambda, \pm} + e^{-\Delta \tau
\omega_\lambda/4}c^\dagger_{\lambda, \mp} \right ].
\end{equation}
This is the familiar Bogoliubov transformation for bosonic field
[22]. Therefore
for the observer, who can only observe the modes
$\varphi_{\lambda, +}$ within the two horizons, the no-boundary
state or vacuum state
$|0_{NB} \rangle$ in (35) will present the Planck spectrum for
radiation at temperature  $\Delta 
\tau^{-1}$. Indeed the observer
will detect
\begin{equation}
\langle 0_{NB}|a_{\lambda, +}^\dagger a_{\lambda, +}
|0_{NB}\rangle =
\frac{1}{e^{\Delta \tau
\omega_\lambda} - 1}
\end{equation}
particles in the mode $\varphi_{\lambda, +}$. To the observer the
quantum state should be described by a density matrix, which
represents a thermal equilibrium state at temperature $\Delta
\tau^{-1}$, as in the Schwarzschild case [17]. If the observer is
in $\Sigma_2$, then the situation will be the same.

For comparison, we show the so-called Boulware vacuum state
observed at the quantum transition surface of the spacetime
background [19]
\begin{equation}
\Psi_{B,\pm} \approx \exp \left [ - \frac{1}{2}
\sum_{\lambda}
\omega_\lambda \varphi_{\lambda, \pm}^2 \right ],
\end{equation}
which are separately defined on $\Sigma_1$ and $\Sigma_2$. The
Boulware modes
are eigenfunctions of the Killing vector $\partial / \partial t$.

Our result is well expected, since the
constrained gravitational instanton with parameter $f_2$ is the
seed of the Schwarzschild-de Sitter spacetime background, the
period of the imaginary time is $\Delta \tau$. It is noticed from
Eq. (15) that the probability of the temperature in the
perturbation modes is constant.

If $m = 0$, our  model is reduced to the de Sitter case, the
black hole horizon disappears and the topology of the
3-geometry becomes $S^3$. The observer can measure the
quantum field in the whole equator $\Sigma$. If one analytically
continues
the basis 
functions (29) into real time at the quantum transition, then one
obtains
\begin{equation}
u_{\lambda, \pm}(t, r, \theta, \phi)
= (\cos \omega t \mp i \coth (\Delta \tau \omega /2) \sin \omega
t ) 
R_{\omega l \bar{m} A }(r, \theta, \phi),
\end{equation}  
since the scalar field operator is Hermitian, the appearance of
the imaginary part of (45) is spurious. One can use the
prescription for the analytic continuation from the imaginary
time to the real time at $\Sigma_1$, that is, to 
set $\tau =  \Delta \tau /4 \pm  i t$ and then take the average.
Then the
imaginary part in (45) is cancelled, the perturbation modes
behave as standing waves, and the dependence of the
later development on the parameter $\Delta \tau$ is therefore
completely erased.

So the choice of
the period in imaginary time used in constructing the instanton
has   effect neither on the background, nor on
the perturbation modes. At the initial moment the quantum state
of the
perturbation modes is strictly at the minimum excitation state
allowed by the Heisenberg Uncertainty Principle, or the
temperature is zero then [20]. If we keep using
the Killing time, then the situation will remain the same
forever. 

However, it is more appropriate to use the so-called synchronous
coordinates in cosmology [10], at least for the background. In
this
coordinate, it would be convenient to decompose the perturbation
in terms of the comoving modes. Then,  the state will remain in
the ground state with a time-dependent frequency which varies
inversely with respect to the scale of the universe, as long as
the wave
length of the mode is smaller than the horizon.  If the mode goes
out the horizon, then its wave function will freeze, it
becomes a superposition of a number of excited states due to
the background evolution.  This
phenomenon leads to the formation of structures in the universe
[20]. The calculation can be carried out more quantitatively by
using the technique of the so-called squeezed vacuum state
developed
in quantum optics [23]. The quantum state observed is described
by a density matrix obtained by taking trace of the wave function
over the field configuration beyond the cosmological horizon.

One can straightforward apply the above argument to the Nariai
case. Since the two
horizons degenerate, no trace has to be taken if one uses the
coordinate with the Killing time. Again, if one uses the
synchronous coordinates, the quantum state will behave in a
similar way as that in the de Sitter model. In summary, the
constrained instanton approach keeps the de Sitter and Nariai
models intact.

In principle, one can also analyze the quantum state in the
Schwarzschild-de Sitter background using the synchronous
coordinates
in a similar way.

\vspace*{0.3in}

\large \bf V. The quantum state of the fermionic field

\vspace*{0.15in}

\rm 
\normalsize

Now we are going to analyze the spin-1/2 fermionic perturbation
modes. The theory of a quantized spin-1/2 field in the
Schwarzschild space [24] will be generalized to the case in the
Schwarzschild-de Sitter space.

The Dirac spinor $\psi$ can be globally defined in the flat
Minkowski space.
However,  in a curved space, one has to locally
introduce a set of orthonormal basis vectors represented by the
tetrad $e^a_\mu$ with respect to which the spinors $\psi$ are
defined. The action of the spinor field is 
\begin{equation}
I_f = \int_M \left ( \frac{i}{2} \tilde{\psi} \gamma
\stackrel{\leftrightarrow}{
\bigtriangledown} \psi  - m_0 \tilde{\psi}\psi \right )
 - \frac{i}{2}
\int_{\partial M}\tilde{\psi}\gamma \psi
\end{equation}
where $m_0$ is the mass of the spinor. In the usual form, one has
$\tilde{\psi} = \psi^\dagger \gamma^0$. Here instead, we shall
free the daggered variables from being the Hermitian conjugates
of undaggered variables by twofold reason.
First, it is due to the variational calculation technique.
Second, it will leave room for the calculation below in
Euclidean approach. The boundary term depends on the variational
condition posed. Like in the purely gravitational case, we
specify the 3-geometry
on the boundary for action (2), here the boundary term is for the
condition that the fermionic variables $\tilde{\psi}$ and
$\psi$ are specified on the  final surface $S_f$ and
initial surface $S_i$  if the two surfaces are
properly defined. 

The Dirac equation can be derived from the action 
\begin{equation}
\left ( -i \gamma^a \bigtriangledown_a + m_0 \right) \psi = 0,
\end{equation}
\begin{equation}
 \tilde{\psi} \left ( i \gamma^a
\stackrel{\leftarrow}{\bigtriangledown}_a + m_0
\right)= 0.
\end{equation}

For the Schwarzschild-de Sitter background, one can choose the
tetrad along the coordinate basis [24] and a proper spinor
representation. Then the Euclidean Dirac
equation for $\psi$ in the sector $r_2 < r < r_3$ can be written
\begin{equation}
\left (  - \frac{\gamma^0 \partial }{\Delta \partial \tau} +
\frac{\gamma^1 \Delta^{1/2} \partial \Delta ^{1/2} r}{ir \partial
r} + \frac{\gamma^2\partial \sin^{1/2} \theta}{ir \sin^{1/2}
\theta \partial \theta} + \frac{\gamma^3 \partial}{ir \sin
\theta \partial \phi } + m_0 \right ) \psi = 0.
\end{equation}
The Dirac equation for $\tilde{\psi}$ can be treated in a similar
way.

One can choose the representation of the Dirac matrices to be a
direct product $\vec{\rho} \otimes \vec{\sigma}$ of two-
dimensional Pauli matrices [24]
\begin{equation}
\gamma^0 = \rho_2, \;\;\gamma^1 = i\rho_1, \;\;\gamma^2 =-i
\rho_3\sigma_3, \;\;\gamma^3 =-i \rho_3\sigma_1,
\end{equation} 
then the basis of solutions of Eq. (49) are
\[
\eta_{\lambda}^\pm
=\exp(\mp \omega \tau)  \psi_{\lambda}^\pm (r, \theta, \phi),  
\]
\begin{equation}
\lambda = ( \omega, k^\prime, m^\prime ),
\end{equation}
where the upper index $\pm$ means the positive (negative)
frequency part. 
Then the eigenfunctions for the spinor field can be written as
\begin{equation}
\psi_{\lambda}^\pm (r, \theta, \phi)  
=r^{-1} \Delta^{-1/2} Y_{k^\prime
m^\prime} (\theta,\phi)\psi_{\lambda}^\pm (r),
\end{equation}
where the radial functions $\psi_{\lambda}^\pm (r)$
live in the $\rho$ subspace, the spinor harmonics
 $Y_{k^\prime m^\prime}$  live in the $\sigma$ subspace, and
$k^\prime$ and $m^\prime$ denote the eigenvalue of
the following operators [21][24]
\begin{equation}
k =   \frac{ \sigma^3\partial 
\sin^{1/2} \theta}{i
\sin^{1/2} \theta \partial \theta}+
\frac{\sigma^1 \partial}{ i\sin \theta \partial
\phi}
\end{equation}
and
\begin{equation}
J_3  = \frac{\partial}{i \partial \phi}.
\end{equation}
For the harmonics with a given total angular momentum $j(j+1)$
the degeneracy is $2(2j +1)$.

The radial function satisfies the equation
\begin{equation}
\left ( \mp \frac{\rho^2 \omega}{\Delta}+ \frac{\rho^1
\partial
}{\partial r^*} - \frac{i\rho^3 k^\prime}{r }+ m_0
\right ) \psi_\lambda^\pm(r)= 0.
\end{equation}

Let us study the spinor field at the surface of the quantum
transition or $\Sigma$. One can
decompose the spinor field as follows
\[
\psi (t, r, \theta, \phi) =\sum_\lambda [m_{\lambda,+}
\eta_{\lambda}^+(it, r_+, \theta, \phi)+
\bar{n}_{\lambda, +}\eta_{\lambda}^-( it, r_+, \theta, \phi)
\]
\begin{equation}
+m_{\lambda,-}
\eta_{\lambda}^+(it, r_-, \theta, \phi)+
\bar{n}_{\lambda, -}\eta_{\lambda}^-( it, r_-, \theta, \phi)],
\end{equation}
\[
\psi^\dagger (t, r, \theta, \phi) =\sum_\lambda
[\bar{m}_{\lambda,+}\bar{\eta}_{\lambda}^+( it, r_+, \theta,
\phi)+
n_{\lambda, +}\bar{\eta}_{\lambda}^-(it, r_+, \theta, \phi)
\]
\begin{equation}
+\bar{m}_{\lambda,-}\bar{\eta}_{\lambda}^+( it, r_-, \theta,
\phi)+
n_{\lambda, -}\bar{\eta}_{\lambda}^-(it, r_-, \theta, \phi)],
\end{equation}
where
\begin{equation}
\sum_\lambda = \frac{1}{\sqrt{2\pi}} \int^\infty_0 d \omega
\sum_{k^\prime,m^\prime }
\end{equation}
and the field $ \psi , \psi^\dagger $ and the
coefficients  $m_\pm, n_\pm, \bar{m}_\pm, \bar{n}_\pm$ 
are taken to be odd elements of a Grassmann algebra. It is
noted that in the expansions (56) (57) we use the Killing time
coordinate $t$. We implicitly assume that the Lorentzian
evolution is along the $t$ increasing (decreasing) direction at
$\Sigma_1 \;\;(\Sigma_2)$.

There is interference neither between different modes nor between
those at $\Sigma_1$ and $\Sigma_2$. Thus it is enough to study
individual
modes separately. The Lagrangian for one mode is
\begin{equation}
L_\lambda = \bar{m}_\lambda(\frac{i \partial}{\partial t}
-\omega_\lambda ) m_\lambda
+\bar{n}_\lambda(\frac{i \partial}{\partial t} +\omega_\lambda )
n_\lambda
\end{equation}
and the Hamiltonian is
\begin{equation}
H_\lambda = \omega_\lambda(\bar{m}_\lambda m_\lambda -
\bar{n}_\lambda n_\lambda).
\end{equation}
The quantum versions of $m_\lambda, n_\lambda, \bar{m}_\lambda ,
\bar{n}_\lambda$ obey the anti-commutation relations
\begin{equation}
\{ m_\lambda , \bar{m}_\lambda \} = 1, \;\;\;\; \{ n_\lambda ,
\bar{n}_\lambda \} = 1.
\end{equation}
In the $m_\lambda , n_\lambda$ representations, one has
\begin{equation}
\bar{m}_\lambda = \frac{\partial}{\partial
m_\lambda},\;\;\;\;\bar{n}_\lambda =
\frac{\partial}{\partial n_\lambda}.
\end{equation}

The Hamiltonian operator can be written as
\begin{equation}
H_\lambda =  \omega_\lambda \left (- m_\lambda
\frac{\partial}{\partial m_\lambda} + n_\lambda
\frac{\partial}{\partial n_\lambda} \right ).
\end{equation}

Within the framework one can define the creation and annihilation
operators for the spinor field at $\Sigma_1$ and $\Sigma_2$
\begin{equation}
a_{\lambda , \pm} =  \frac{\partial}{\partial 
m_{\lambda, \pm}}, \;\;\; a_{\lambda , \pm}^\dagger = m_{\lambda,
\pm},
\end{equation}
\begin{equation}
b_{\lambda , \pm} = n_{\lambda, \pm},  \;\;\; b_{\lambda ,
\pm}^\dagger
=\frac{\partial}{\partial n_{\lambda, \pm}}. 
\end{equation}

Then the vacuum state of the spinor field with respect to these
operators, or the wave function of the minimum excitation  state
for the field defined only at $\Sigma_1 \;(\Sigma_2)$, is
\begin{equation}
\Psi_{B, \pm}(h_{ij},\psi) = n_{\lambda, \pm}.
\end{equation}

Now we are going to evaluate the no-boundary wave function for
the fermionic perturbation modes. At the
$WKB$ level, the wave function is the exponential of the negative
of the Euclidean action for the
classical solution. The action (46) is suitable for the solution
with the boundary condition that the
fermionic variables $\tilde{\psi} $ and $\psi$ are specified at
the final and initial surfaces, or $\Sigma_1$ and $\Sigma_2$
respectively.

The dominating contribution to the no-boundary wave function is
due to the Euclidean solution satisfying the above boundary
condition. Since $\Sigma$ is also the surface of quantum
transition, at which or the moment $t = 0$ the Lorentzian
evolution will emanate, the spinor solution in the Lorentzian
regime can be obtained through an analytic continuation from that
in the Euclidean regime and it can be written as a linear
combination
\[
\psi (t, r, \theta, \phi) =\sum_\lambda [m_{\lambda,-}
\eta_{\lambda}^+(it, r_+, \theta, \phi)\exp (-\Delta \tau
\omega_\lambda/2)+n_{\lambda, -}
\eta_{\lambda}^-( it, r_+, \theta, \phi)\exp
(\Delta \tau \omega_\lambda/2)
\]
\begin{equation}
 +m_{\lambda,-}\eta_{\lambda}^+(- it, r_-, \theta, \phi)+
n_{\lambda, -}
\eta_{\lambda}^-(-it, r_-, \theta, \phi)],
\end{equation}
\[
\tilde{\psi} (t, r, \theta, \phi) =\sum_\lambda
[m_{\lambda,+} \bar{\eta}_{\lambda}^+( it, r_+, \theta, \phi)+
n_{\lambda, +}\bar{\eta}_{\lambda}^- (it, r_+, \theta, \phi)
\]
\begin{equation}
 +m_{\lambda,+}
\bar{\eta}_{\lambda}^+(- it, r_-, \theta, \phi)\exp
(-\Delta \tau \omega_\lambda/2)+n_{\lambda, +}
\bar{\eta}_{\lambda}^-(-it, r_-, \theta, \phi)\exp
(\Delta \tau \omega_\lambda/2)].
\end{equation}
To comply with the boundary term of the action (46), the
coefficients
$m_{\lambda,+}, n_{\lambda,+} (m_{\lambda, -}, n_{\lambda,-})$
are determined by the decomposition of
the field at the final and initial surfaces, or $\Sigma_1$ and
$\Sigma_2$ respectively. 

The Euclidean action  can be evaluated by substituting (67)
and (68) into (46). For the classical solution, only the boundary
term of the action survives,
\begin{equation}
\bar{I}_f=  \sum_\lambda [ m_{\lambda,+}m_{\lambda,-}\exp
(-\Delta \tau \omega_\lambda /2) +n_{\lambda, +}n_{\lambda,
-}\exp (\Delta \tau \omega_\lambda /2)].
\end{equation}
The no-boundary wave function for the fermionic perturbations can
be written as a product of that for each mode. Therefore, the
no-boundary wave function of mode $\lambda$ is
\[
\Psi_{NB}(h_{ij},m_{\lambda,+},m_{\lambda,-},n_{\lambda,
+},n_{\lambda,-}) \approx \exp -\bar{I}_{f,\lambda }
\]
\begin{equation}
= 1 -m_{\lambda,+}m_{\lambda,-}\exp (
-\Delta \tau \omega_\lambda /2) -n_{\lambda, +}n_{\lambda,
-}\exp (\Delta \tau \omega_\lambda /2)   +
m_{\lambda,+}m_{\lambda,-}n_{\lambda, +}n_{\lambda,-}.
\end{equation}

To interpret (70), it is convenient to introduce the operators
\begin{equation}
c_{\lambda , \pm} =\left ( 2\cosh \frac{\Delta \tau
\omega_\lambda}{2} \right )^{-1/2} \left [e^{\Delta \tau
\omega_\lambda/4}\frac{\partial}{\partial m_{\lambda, \pm}} \pm
e^{-\Delta \tau \omega_\lambda/4}m_{\lambda, \mp} \right ],
\end{equation}
\begin{equation}
d_{\lambda , \pm} =\left (2\cosh \frac{\Delta \tau
\omega_\lambda}{2} \right )^{-1/2} \left [e^{\Delta \tau
\omega_\lambda/4} n_{\lambda, \pm}\mp
e^{-\Delta \tau \omega_\lambda/4}\frac{\partial}{\partial
n_{\lambda, \mp}} \right ]
\end{equation}
and their adjoint
\begin{equation}
c_{\lambda , \pm}^\dagger =\left (2\cosh \frac{\Delta \tau
\omega_\lambda}{2} \right )^{-1/2} \left [e^{\Delta \tau
\omega_\lambda/4} m_{\lambda, \pm} \pm e^{-\Delta \tau
\omega_\lambda/4}\frac{\partial}{\partial m_{\lambda, \mp}}
\right ],
\end{equation}
\begin{equation}
d_{\lambda , \pm}^\dagger =\left ( 2\cosh \frac{\Delta \tau
\omega_\lambda}{2} \right )^{-1/2} \left [e^{\Delta \tau
\omega_\lambda/4}\frac{\partial}{\partial  n_{\lambda, \pm}} \mp
e^{-\Delta \tau \omega_\lambda/4} n_{\lambda, \mp} \right ].
\end{equation}

These operators are the creation and annihilation
operators which satisfy the corresponding anti-commutative
relations. It is noted that if one sets $\Delta \tau = \infty$,
then Eqs. (71) to (74)  will be reduced to Eqs. (64)(65).
One can verify that the no-boundary wave function (70)
is the ground state with respect to these operators. That is, one
has
\begin{equation}
c_{\lambda, \pm} \Psi_{NB} (h_{ij}, \psi ) = 0
\end{equation}
and
\begin{equation}
d_{\lambda, \pm} \Psi_{NB} (h_{ij}, \psi ) = 0.
\end{equation}

Therefore, if there exists an omniscient being who can perceive
the quantum state on the whole surface of quantum transition, he
would find that the spinor modes are in a pure state, and even
more than this, that is the state is in vacuum. However, the
reality is that one can only observe the quantum state within his
black hole and cosmological horizons, say at $\Sigma_1$. Since
the information beyond these horizons is lost,
he would find that the quantum state is described by a
density matrix. In fact, one can even expect that it is in a
thermal equilibrium state, as we shall show below.

From Eqs. (64)-(65)(71)-(74) one can derive the following
relations:
\begin{equation}
a_{\lambda , \pm} =\left ( 2\cosh \frac{\Delta \tau
\omega_\lambda}{2} \right )^{-1/2} \left [e^{\Delta \tau
\omega_\lambda/4} c_{\lambda, \pm} \mp
e^{-\Delta \tau \omega_\lambda/4}c^\dagger_{\lambda, \mp} \right
]
\end{equation}
and
\begin{equation}
b_{\lambda , \pm} =\left ( 2\cosh \frac{\Delta \tau
\omega_\lambda}{2} \right )^{-1/2} \left [e^{\Delta \tau
\omega_\lambda/4}d_{\lambda, \pm}\pm e^{-\Delta \tau
\omega_\lambda/4}d_{\lambda, \mp}^\dagger \right ].
\end{equation}

These are the  Bogoliubov transformations for fermionic
fields. The sign differences are due to the fact that $m\;\; (n)$
modes are associated with the positive (negative) frequency part
of the solution. Since the direction of Killing time vector
flips at surfaces $\Sigma_1$ and $\Sigma_2$, there exists a
duality of creation and annihilation processes between surfaces
$\Sigma_1$ and $\Sigma_2$. Therefore, the vacuum state observed
by the omniscient being appears as a thermal
equilibrium state for the ordinary observer like us who is
restricted by the two horizons. The associated temperature is the
inverse of $\Delta \tau$ as shown by (77)-(78). Indeed the
observer will detect
\begin{equation}
\langle 0_{NB}|a_{\lambda, +}^\dagger a_{\lambda, +}
|0_{NB}\rangle =
\frac{1}{e^{\Delta \tau
\omega_\lambda} + 1}
\end{equation}
particles in $m_{\lambda, +}$ mode and
\begin{equation}
\langle 0_{NB}|b_{\lambda, +}^\dagger b_{\lambda, +}
|0_{NB}\rangle =
\frac{1}{e^{\Delta \tau
\omega_\lambda} + 1}
\end{equation}
particles in $n_{\lambda, +}$ mode. If the observer is in
$\Sigma_2$, then the situation will be the same.

The alternative
way to show this thermal property of the state observed at
$\Sigma_1\;\; (\Sigma_2)$ is to reduce the wave function into a
density matrix by taking the trace over the unobserved
configuration at $\Sigma_2\;\; (\Sigma_1)$  [17].

The argument about the two special cases, the de Sitter and the
Nariai models in the last section remains intact for the spinor
field. If one sets $m=0$, then the model is reduced to the case
of
the  spinor field in the de Sitter background [25].

\vspace*{0.3in}

\large \bf VI. Discussion
\vspace*{0.15in}

\rm 
\normalsize

The possibility of quantum  creation of a single black hole in
the de Sitter background has been argued for a long period of
time. The reason is that there does not exist any
regular gravitational instanton which is the seed of the created
hole. However, in quantum theory, one is mainly concerned about
the stationary action solution. If one can find a stationary
action solution, then the wave function can be approximated by a
classical trajectory. The stationary action solutions are not
always the solutions to the field equation. The set of all
solutions to the classical field equations is only a subset of
the class of all stationary action solutions. In this sense, we
introduce the concept of constrained instanton. In quantum
cosmology we are particularly interested in the
constrained instantons for which the 3-geometry of the quantum
transition surface is given.

We find that there exists a free parameter in the constrained
gravitational instanton for the black hole creation in the de
Sitter background. This is the identification
period of imaginary time in constructing the instanton. It is
naturally expected that the inverse of this parameter should be
the temperature of quantum fields in the spacetime background. We
studied both bosonic and fermionic perturbation modes and
confirmed that this was indeed the case.

It was thought that in the Schwarzschild-de Sitter space, the
quantum
state of the perturbation could never be in thermal equilibrium,
since the temperatures associated  with the two horizons are
different. It turns out that Nature is more generous than we
expected.  The quantum fields can coexist with the black hole
spacetime in an equilibrium state at an arbitrary temperature. 

When we discussed the quantum state of the perturbation modes, we
treated spacetime as a classical background. We ignored both the
interaction between the gravitational and the matter field modes
and the back reaction of the perturbation modes to the
background. We assumed that their effects were negligible.

Quantum fields in curved spacetime is an incomplete and
temporary theory. Practically, it can only be useful for some
very symmetric backgrounds. Conceptually, one can
foresee a very dramatic revolution in this direction in the not
so distant future.
\vspace*{0.3in}

\bf References:

\vspace*{0.1in}
\rm

1. J.B. Hartle and S.W. Hawking, \it Phys. Rev. \rm \bf D\rm
\underline{28}, 2960 (1983).

2. S.W. Hawking, \it Nucl. Phys. \rm \bf B\rm \underline{239},
257 (1984).

3. S.W. Hawking, \it Commun. Math. Phys. \rm \underline{43}, 199
(1975).

4. G.W. Gibbons and S.W. Hawking, \it Phys. Rev. \bf D\rm
\underline{15}, 2738 (1977).

5. R. Bousso and S.W. Hawking, \it Phys. Rev. \rm \bf D\rm
\underline{52}, 5659 (1995).

6. Z.C. Wu, \it Int. J. Mod. Phys. \rm \bf D\rm\underline{6}, 199
(1997).

7. G. Hayward and J. Louko, \it Phys. Rev. \rm \bf D\rm
\underline{42}, 4032 (1990).

8. G.W. Gibbons and J.B. Hartle, \it Phys. Rev. \rm \bf D\rm
\underline{42}, 2458 (1990).

9. X.M. Hu and Z.C. Wu, \it Phys. Lett. \rm \bf B\rm
\underline{149}, 87 (1984).

10. Z.C. Wu,  \it Prog. Theo. Phys. \rm 
\underline{97}, 859 (1997);  \it Prog. Theo. Phys. \rm 
\underline{97}, 873 (1997).

11. R.B. Mann and S.F. Ross, \it Phys. Rev. \rm \bf D\rm
\underline{52}, 2254 (1995).

12. S.F. Ross and S.W. Hawking, \it Phys. Rev. \rm \bf D\rm
\underline{52}, 5865 (1995).

13. Z.C. Wu,  (1998)(to appear).

14. S.W. Hawking, \it Commun. Math. Phys.
\rm \underline{43}, 199 (1975).

15. J.B. Hartle and S.W. Hawking, \it Phys. Rev.
\rm \bf D\rm \underline{13}, 2188 (1976).

16. G.W. Gibbons and M.J. Perry, \it Proc. Roy. Soc. London \rm
\bf A\rm \underline{358}, 467 (1989).

17. R. Laflamme, \it Nucl. Phys. \rm \bf B\rm \underline{324},
233 (1978).

18. A.O. Barvinsky, V.P. Frolov and A.I. Zelnikov, \it Phys. Rev.
\rm \bf D\rm \underline{51}, 1741 (1995).

19. D.G. Boulware, \it Phys. Rev. \rm \bf D\rm \underline{11},
1404 (1975).

20. J.J. Halliwell and S.W. Hawking, \it Phys. Rev. \rm \bf D\rm
\underline{31}, 1777 (1985).

21. P.A.M. Dirac, \it The Principle of Quantum Mechanics, \rm
(Oxford Clarendon Press) (1938).

22. N.D. Birrell and P.C.W. Davies, \it Quantum Fields in Curved
 Space, \rm ( Cambridge University Press) (1982).

23. L.P. Grishchuk and Y.V. Sidorov, \it Class. Quantum Grav. \rm
\underline{6}, \bf L\rm 161 (1989).

24. D.G. Boulware, \it Phys. Rev. \rm \bf D\rm \underline{12},
350 (1975).

25. P.D. D'Eath and J.J. Halliwell, \it Phys. Rev. \rm \bf D\rm
\underline{35}, 1100 (1987).

\newpage

\end{document}